\documentstyle[epsfig]{mn2e}

\def\be{\begin{equation}}
\def\ee{\end{equation}}
\def\bea{\begin{eqnarray}}
\def\eea{\end{eqnarray}}

\title[]{GRB efficiency and Possible Physical Processes Shaping the Early Afterglow}
\author[]{Yizhong Fan$^{1,2,3}$\thanks{Lady Davis Fellow, E-mail: yzfan@pmo.ac.cn} and Tsvi Piran$^1$
\thanks{tsvi@phys.huji.ac.il}\\
$^1${\sl The Racah Inst. of Physics, Hebrew University, Jerusalem 91904, Israel}\\
$^2$ {\sl Purple Mountain Observatory, Chinese Academy of
Science, Nanjing 210008, China}\\
$^3${\sl National Astronomical Observatories, Chinese Academy of
Sciences, Beijing 100012, China}\\}
\date{Accepted ......
Received ......; in original form ......}

\begin{document}

\maketitle
\begin{abstract}
The discovery by {\em Swift} that a good fraction of Gamma Ray
Bursts (GRBs) have a slowly decaying X-ray afterglow phase led to
the suggestion that energy injection into the blast wave takes
place several hundred seconds after the burst. This implies that
right after the burst the kinetic energy of the blast wave was
very low and in turn the efficiency of production of $\gamma$-rays
during the burst was extremely high, rendering the internal shocks
model unlikely. We re-examine the estimates of kinetic energy in
GRB afterglows and show that the efficiency of converting the
kinetic energy into $\gamma-$rays is moderate and does not
challenge the standard internal shock model. We also examine
several models, including in particular energy injection,
suggested to interpret this slow decay phase.  We show that with
proper parameters, all these models give rise to a slow decline
lasting several hours. However, even those models that fit all
X-ray observations, and in particular the energy injection model,
cannot account self-consistently for both the X-ray and the
optical afterglows of well monitored GRBs such as GRB 050319 and
GRB 050401. We speculate about a possible alternative resolution
of this puzzle.
\end{abstract}

\begin{keywords}
Gamma Rays: bursts$-$ISM: jets and outflows--radiation
mechanisms: nonthermal$-$X-rays: general
\end{keywords}


\section{Introduction}
\label{sec:XRF1} The X-ray telescope (XRT) on board {\em Swift}
has provided high quality early X-ray afterglow light curves of
many Gamma-ray Bursts (GRBs). One of the most remarkable and
unexpected features discovered by {\em Swift} was that many of
these X-ray afterglow light curves are distinguished by a slow
decline---The flux $F$ decreases with observer's time $t$ as
$F\propto t^{[0,-0.8]}$, lasting from a few hundred seconds to few
hours (Nousek et al. 2005, Campana et al. 2005; Vaughan et al.
2005; Cusumano et al. 2005; de Pasquale et al. 2005).  Such a
phase is unexpected in the standard fireball model. A simple
explanation is that the slow decline arises due to a significant
energy injection (Zhang et al. 2005; Nousek et al. 2005;
Panaitescu et al. 2006, Granot \& Kumar 2006), as suggested
previously (For baryon-rich injection: Rees \& M\'esz\'aros 1998;
Panaitescu et al. 1998; Kumar \& Piran 2000; Sari \& M\'esz\'aros
2000; Zhang \& M\'esz\'aros 2002; Granot, Nakar \& Piran 2003. For
Poynting flux dominated injection\footnote{If the outflow ejected
from the central engine after the gamma ray burst phase is highly
magnetized, at a radius $\sim 10^{15}$ cm, the MHD condition
breaks down. Significant magnetic field dissipation processes are
expected to happen which converts energy into radiation. As long
as the highly magnetized outflow is steady enough, strong and
slowly decaying X-ray emission is possible (see Fan, Zhang \&
Proga [2005a] and the references therein).}: Dai \& Lu 1998a;
Zhang \& M\'esz\'aros 2001; Dai 2004). It has been argued that
consequently the resulted GRB efficiency, i.e., the ratio of the
energy emitted in  $\gamma-$ray energy to the total energy (the
sum of the $\gamma-$ray energy and the kinetic energy of the
ejecta powering the afterglow), should be $90\%$ or higher. Some
extreme assumptions are needed (Beloborodov 2000; Kobayashi \&
Sari 2001) to reach such a high efficiency within the framework of
the standard internal-shocks model (Paczynski \& Xu 1994; Rees \&
M\'esz\'aros 1994; Sari \& Piran 1997a , 1997b; Kobayashi, Piran
\& Sari 1997; Daigne  \& Mochkovitch 1998; Piran 1999).

We re-examine this issue focusing on two critical aspects of the
analysis. The estimate of the kinetic energy of the ejecta from
the afterglow observations and in particular from the X-ray flux
and the need of energy injection.  We show in \S\ref{sec:eff} that
even for these {\em Swift} GRBs with long duration X-ray
flattening the $\gamma$-ray conversion efficiency is high but not
unreasonable.

We then turn to the puzzling slow decline seen in the first few
hours of the X-ray afterglow. We explore in  \S\ref{sec:XRF2}
several models that may give rise to slowly decaying X-ray
afterglows: (i) Energy injection. (ii) A small $\zeta_e$, in which
only a small fraction, $\zeta_e \ll 1$ of the electrons are
accelerated to high energies and contribute to the radiation
process. (iii) Evolving shock parameters, where the microscopic
shock parameters $\epsilon_e$ and/or $\epsilon_B$ (the fraction of
shock energy given to the magnetic filed) vary in time and are
inversely proportional to the Lorentz factor of the ejecta. (iv) A
very low variable external density model, in which the number
density of the medium is not only very low but it also a function
of the radius. (v) Highly magnetized outflow where flattening
might arise because of a slow conversion of the magnetic energy to
kinetic energy of the external matter. We present in
\S\ref{sec:XRF2} analytical derivation as well as numerical
calculations of the expected light curves in all these models
except the last one. In \S\ref{sec:0319} we compare the models to
the observations of GRB 050319 and  GRB 050401. We summarize our
results and discuss their implications in \S\ref{sec:XRF4}. We
conclude with a speculation on the nature of the solution to this
puzzle.


\section{Is there a GRB efficiency crisis?}
\label{sec:eff}

One of the critical factors that characterize the emitting of a
GRB is the energy conversion efficiency. The $\gamma$-ray
efficiency is defined as:
\begin{equation}
\label{eq:eff} \epsilon_\gamma\equiv E_\gamma/(E_\gamma+E_k) \ ,
\end{equation}
where $E_\gamma$ is the isotropic equivalent energy of the
$\gamma-$ray emission and $E_k$ is the isotropic equivalent energy
of the outflow powering the afterglow. Following the {\em Swift}
observations of flattening in the X-ray afterglow light curve of
many GRBs,  it has been argued that  typical values of
$\epsilon_\gamma$ could be as high as 90\% or even higher (Zhang
et al. 2005; Nousek et al. 2005; Ioka et al. 2005; for the
discussion of pre-{\em Swift} GRBs, see Llod-Ronning \& Zhang
2004, hereafter LZ04). This very high efficiency would challenge
most $\gamma$-rays emission models and in particular it challenges
the standard fireball model that is based on internal shocks.

These claims arise from revised estimates of the kinetic energy
immediately following the GRB.  Therefore, in order to explore
this issue we re-examine the estimates of the kinetic energy from
the X-ray observations. As we show below, at a late afterglow
epoch, the X-ray band is above the cooling frequency. In this case
the X-ray flux is independent of the poorly constrained $n$ and
the X-ray luminosity is a good probe of $E_k$ (Freedman \& Waxman
2001, Kumar 2000, LZ04).

In the standard GRB afterglow model (e.g., Sari, Piran \& Narayan
1998; Piran 1999), the X-ray afterglow is produced by a shock
propagating into the circum-burst matter. The equations that
govern the emission of this shock  are (Yost et al.
2003)\footnote{To derive these equations, the deceleration of the
fireball is governed by the energy conservation $\Gamma^2 M
c^2=E_k$, where $M$ is the rest mass of the shocked medium (e.g.,
Blandford \& McKee 1976; Sari et al. 1998; Piran 1999). The
distribution of the fresh electrons accelerated by the shock is
assumed to be $dn/d\gamma_e \propto \gamma_e^{-p}$ for
$\gamma_e\geq \gamma_{e,m}$, where
$\gamma_{e,m}=(m_p/m_e)[(p-2)/(p-1)]\epsilon_e(\Gamma-1)$,
governed by the strict shock jump conditions (Blandford \& McKee
1976). The other crucial parameter is the cooling Lorentz factor
$\gamma_{e,c}=6(1+z)\pi m_e c/[\sigma_T \Gamma B^2 t(1+Y)]$, above
which the energy loss due to the synchrotron/inverse-Compoton
radiation is important (Sari et al. 1998; Piran 1999), where
$\sigma_T$ is the Thompson cross section and $B$ is the magnetic
field of the shocked medium. $\nu_m$ and $\nu_c$ are the
corresponding synchrotron radiation frequency of electrons with
Lorentz factor $\gamma_{e,m}$ and $\gamma_{e,c}$, respectively.
The maximum specific flux is estimated as  $F_{\nu, \rm
max}\approx (1+z)M \Gamma e^3 B/(4\pi m_p m_e c^2 D_L^2)$ (Sari et
al. 1998; Wijers \& Galama 1999), where $e$ is the charge of
electron. The $\nu_c$ and $F_{\nu, \rm max}$ taken here are
comparable with that of most previous works (e.g., Granot et al
1999; Wijers \& Galama 1999; Panaitescu \& Kumar 2002; LZ04). The
$\nu_m$ is close to that taken in Sari et al. (1998), Granot et
al. (1999) and Wijers \& Galama (1999), but is about $30-40$ times
smaller than that taken in Panaitescu \& Kumar (2002) and LZ04.
Such a large divergency may arise if one ignores the term
$(p-2)/(p-1)$ when evaluating $\gamma_{e,m}$.}
\begin{equation}  F_{\nu,{\rm
max}}=6.6~{\rm mJy}~({1+z\over 2}) D_{L,28.34}^{-2}
\epsilon_{B,-2}^{1/2}E_{k,53}n_0^{1/2},\label{eq:F_max}
\end{equation}
\begin{equation} \nu_m = 7.6\times 10^{11}~{\rm Hz}~E_{\rm
k,53}^{1/2}\epsilon_{\rm B,-2}^{1/2}\epsilon_{e,-1}^2 C_p^2 ({1+z
\over 2})^{1/2} t_d^{-3/2},\label{eq:nu_m}
\end{equation}
\begin{equation}
\nu_c = 1.4\times 10^{15}~{\rm Hz}~E_{k,
53}^{-1/2}\epsilon_{B,-2}^{-3/2}n_0^{-1}
 ({1+z \over 2})^{-1/2}t_d^{-1/2}{1\over (1+Y)^2}\label{eq:nu_c},
\end{equation}
where $z$ is the redshift, $D_L$ is the corresponding luminosity
distance, $p$ is the power-law index of the shocked electrons,  we
use $p=2.3$ throughout this work, $C_p \equiv 13(p-2)/[3(p-1)]$
and $t_d$ is the observer's time in unit of days.
$Y=(-1+\sqrt{1+4\eta \eta_{_{KN}} \epsilon_e/\epsilon_B})/2$ is
the Compoton parameter, where
$\eta=\min\{1,(\nu_m/\nu_c)^{(p-2)/2}\}$ (e.g. Sari, Narayan \&
Piran 1996; Wei \& Lu 1998, 2000), $0\leq \eta_{_{KN}}\leq 1$ is a
coefficient accounting for the Klein-Nishina effect, which is
$\gamma_e$ (the random Lorentz factor of the electron) dependent
(see Appendix \ref{sec:Y} for detail). Here and throughout this
text, the convention $Q_x=Q/10^x$ has been adopted in cgs units.

For the typical parameters taken here, $\nu_m$ crosses the
observer frequency $\nu_X\sim 10^{17}$ Hz at $t_d \sim 4\times
10^{-4}$. It is quite reasonable to assume $\nu_X>\max\{ \nu_c,
\nu_m\}$, and the predicted X$-$ray flux is
\begin{eqnarray}
F_{\nu_X} &=& F_{\rm \nu,max}\nu_c^{1/2}\nu_m^{\rm (p-1)/2}
\nu_X^{\rm -p/2}\nonumber\\
&=& 3.8\times 10^{-4}~{\rm mJy}~({1+z \over 2})^{\rm
(2+p)/4}D_{\rm L,28.34}^{-2}\epsilon_{B,-2}^{\rm (p-2)/4}
\epsilon_{\rm e,-1}^{\rm p-1} \nonumber\\
&& E_{k, 53}^{\rm (p+2)/4}(1+Y)^{-1}t_d^{\rm (2-3p)/4}
\nu_{X,17}^{-p/2}.
\end{eqnarray}
The flux recorded by XRT is
\begin{eqnarray}
{\cal{F}} &=& \int^{\nu_{X2}}_{\nu_{X1}} F_{\nu_X} d \nu_X
\nonumber\\ &=& 1.2\times 10^{-12}~{\rm ergs
~s^{-1}~cm^{-2}}~({1+z \over 2})^{\rm (p+2)/4} D_{L,28.34}^{-2}
\nonumber\\
&& \epsilon_{B,-2}^{\rm (p-2)/4} \epsilon_{e,-1}^{\rm
p-1}E_{k,53}^{\rm (p+2)/4}(1+Y)^{-1}t_d^{\rm (2-3p)/4},
\label{eq:F}
\end{eqnarray}
where $\nu_{X1}=0.2$ keV and $\nu_{X2}=10$ keV. This equation is
now inverted to obtain $E_k$ from the observed flux.

In some special cases, $\nu_m<\nu_X<\nu_c$, the flux recorded by
XRT should be
\begin{eqnarray}
{\cal{F}} &=& 1.5\times 10^{-11}~{\rm ergs ~s^{-1}~cm^{-2}}~({1+z
\over 2})^{\rm (p+3)/4} D_{L,28.34}^{-2}
\nonumber\\
&& \epsilon_{B,-2}^{\rm (p+1)/4} \epsilon_{e,-1}^{\rm
p-1}n_0^{1/2}E_{k,53}^{\rm (p+3)/4}t_d^{\rm 3(1-p)/4},
\label{eq:F2}
\end{eqnarray}

\subsection{The efficiency of the pre-{\em Swift} GRBs}
With equation (\ref{eq:F}), the corresponding X-ray luminosity at
$t_d=0.4$ ($\sim 10$h, to compare with the results of LZ04) is
\begin{eqnarray}
L_X &=& 4\pi D_L^2{\cal{F}}/(1+z)\nonumber\\
&=& 1.1\times 10^{46}~{\rm ergs~s^{-1}~cm^{-2}}~({1+z \over
2})^{\rm (p-2)/4} \nonumber\\
&& \epsilon_{\rm B,-2}^{\rm (p-2)/4} \epsilon_{\rm e,-1}^{\rm p-1}
(1+Y)^{-1}E_{k,53}^{\rm (p+2)/4}, \label{eq:L_X}
\end{eqnarray}
which in turn yields
\begin{eqnarray}
E_k &= & 9.2\times 10^{52}~{\rm ergs}~ R ~L_{X,46}^{\rm 4/(p+2)}
({1+z
\over 2})^{\rm (2-p)/(p+2)}\nonumber\\
&& \epsilon_{B,-2}^{\rm -(p-2)/(p+2)}\epsilon_{\rm e,-1}^{\rm
4(1-p)/(p+2)}(1+Y)^{4/(p+2)}, \label{eq:E_iso}
\end{eqnarray}
where $R \sim [t(10h)/T_{90}]^{17\epsilon_e/16}$ is a factor
accounting for the energy radiative loss during the first 10 hours
following the prompt gamma-ray emission phase (Sari 1997; LZ04),
$T_{90}$ is the duration of the GRB. The numerical factor of our
equation (\ref{eq:E_iso}) is larger than that of equation (7) of
LZ04 by a factor of $9.2(1+Y)^{4/(p+2)}$ due to the facts that (1)
The $\nu_m$ taken here, which matches the numerical result better
(one can verify this with a simple code to calculate the dynamical
evolution as well as $\nu_m$ numerically), is about one and half
orders smaller than that taken in LZ04. (2) The inverse Compton
effect has been taken into account. Similar conclusions have been
reached by Granot, K\"{o}nigl \& Piran (2006). However, it is not
easy to estimate $Y$ since it depends on $\epsilon_B$ sensitively
(see Appendix \ref{sec:Time} for discussion). One good way to
estimate the GRB efficiency may be to take $Y\sim 0$ and
$(1+Y)\sim (\epsilon_e/\epsilon_B)^{1/2}$, respectively. In both
cases, our estimates of $\epsilon_\gamma$ (Table \ref{tab:1}) are
significantly lower than those of LZ04.\footnote{While our results
are very close to the recent calculations of Granot, K\"{o}nigl \&
Piran (2006), they also show that the estimates of $E_k$ are very
sensitive to the exact expressions used for  $\nu_c$, $\nu_m$, and
$F_{\rm \nu,max}$. Similar conclusion can be drawn by comparing
previous results of Granot et al (1999), Wijers \& Galama (1999),
Freedman \& Waxman (2001), Panaitescu \& Kumar (2002) and LZ04.
Therefore, an alternative explanation for the apparent high
efficiencies is that the blast wave energy estimates using $L_X$
are simply inaccurate.} Smaller $\epsilon_\gamma$ may be possible
in view of that both $\epsilon_e$ and $\epsilon_B$ might be
significantly lower than the standard parameters taken here
(Panaitescu \& Kumar 2002). {\em We suggest that the typical GRB
efficiency of these pre-Swift bursts is $\sim 0.1$ (see Table
\ref{tab:1} for detail). Such values are well understood within
the internal shock model.}

Additional support for this conclusion arises from late energy
estimates. Berger, Kulkarni \& Frail (2004) used the late time
radio observation to estimate the  kinetic energy at this stage.
The find high energies and correspondingly low $\gamma$-ray
efficiency. For GRB 970508 and GRB 970803, the efficiencies are
0.03 and 0.2 respectively, which coincide with our estimates (see
Table \ref{tab:1}).


\begin{table*}
  \begin{center}
    \caption{\label{tab:1} \small GRB energies and efficiencies,
   $L_X$ used in equation (\ref{eq:E_iso}) and $E_\gamma$
    are all taken from LZ04. The numerical values quoted in
    parentheses are for $(1+Y)\simeq (\epsilon_e/\epsilon_B)^{1/2}$.}
    \begin{minipage}{18.5cm}
      \begin{tabular}{cccccc} \hline\hline
        GRB  & $E_\gamma/10^{52}{\rm ergs}$ & $E_k/10^{52}{\rm ergs}$ &
        efficiency $\epsilon_\gamma$
         \\ \hline

970228  & 1.42 & 17.5 (47.5) &
 0.08 (0.03) \\
970508  & 0.55 & 9.1 (24.8) &
 0.06 (0.02) \\
970828  & 21.98 & 37.4 (101.5) &
 0.37 (0.18) \\
971214 & 21.05 &   78.0 (212) & 0.21 (0.09) \\ 980613 & 0.54 &  11.2 (30.5)  &  0.05 (0.02)\\
980703 & 6.01 & 22.2 (60.2) & 0.21 (0.09) \\ 990123 & 143.8 &  186.6 (507) & 0.43 (0.22)\\
990510 & 17.6 &  121.1 (329) & 0.13 (0.05) \\
990705 & 25.6 & 3.1 (8.5) & 0.89 (0.75) \\ 991216 & 53.5 & 337.1 (916) & 0.14 (0.06) \\
000216 & 16.9 & 4.6 (12.5) &  0.78 (0.58) \\
000926 & 27.97 & 91.7 (249.3) & 0.23 (0.1) \\ 010222 & 85.78 & 209.7 (569.8) & 0.29 (0.13) \\
011211 & 6.72 &  12.1 (33) & 0.36 (0.17)\\
020405 & 7.2 & 42.3 (115) & 0.15 (0.06) \\ 020813 & 77.5 & 203.9 (554) & 0.28 (0.12) \\
021004 & 5.56 & 76.8 (208.8) & 0.07 (0.03)\\
XRF 020903 & 0.0011 & 0.09 (0.25) & 0.01 (0.004)\\
\hline
      \end{tabular}

\vspace*{-0.4cm}
    \end{minipage}
  \end{center}
\end{table*}

\begin{figure}
\begin{picture}(0,200)
\put(0,0){\includegraphics{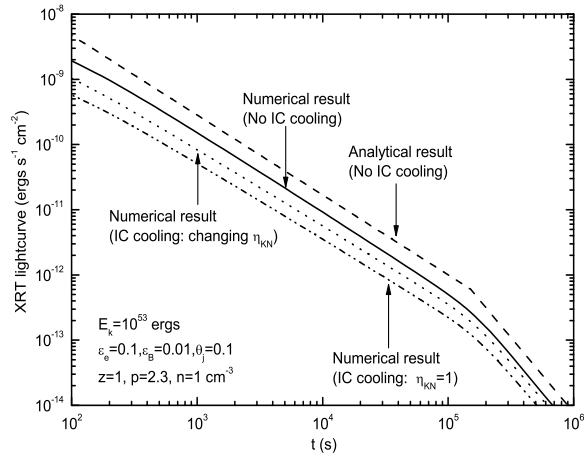}}
\end{picture}
\caption{X-ray (0.2-10 keV) afterglow light curves: Analytical
(dashed line) lightcurve, and numerical (solid line) when Inverse
Compoton effect has been ignored. The divergence is about a factor
of 2. Numerical estimates when the inverse Compoton effect has
been taken into account with (dotted line) and without (dashed -
doted line) a Klein-Nishina correction.  Clearly,  the
Klein-Nishina correction is unimportant for the fiducial
parameters listed in the figure.} \label{fig:eff1}
\end{figure}

The coefficient of our equation (\ref{eq:E_iso}) are very
different from that of equation (7) of LZ04. Below we check its
validity numerically. The code used here has already been used in
Zhang et al. (2005) and has been tested by J. Dyks independently
(Dyks, Zhang \& Fan 2005). Here we just describe  briefly the
technical treatment. The dynamical evolution of the outflow is
calculated with the formulae presented in Huang et al. (2000),
which are able to describe the dynamical evolution of the outflow
in both the relativistic and the non-relativistic phases.  The
electron energy distribution is calculated by solving the
continuity equation with the power-law source function
$Q=K\gamma_e^{-p}$, normalized by a local injection rate. The
cooling of the electrons due to both synchrotron and inverse
Compton  (Moderski, Sikora \& Bulik 2000) has been taken into
account.

Fig. \ref{fig:eff1} depicts the numerical results. One can see
that the numerical results match the analytical ones to within  a
factor of 2. We therefore conclude that equation (\ref{eq:F}) and
equation (\ref{eq:E_iso}) are reasonable approximations to the
full solution of the problem.

\subsection{The GRB efficiency of {\em Swift} GRBs with X-ray flattening}
Early flattening is evident for a good fraction of the X-ray
afterglow light curves recorded by the {\em Swift} XRT.
Determination of the  GRB efficiency of these GRBs is quite
challenging since, as we see in \S\ref{sec:0319}  the underlying
physical process that controls the slow decline is unclear. A
common interpretation for this flat decay is energy injection,
which essentially increases the required initial GRB efficiency.
In spite of the uncertainties concerning the  applicability of
this model we consider its implication to the efficiency.

The energy injection is characterized by a factor $f$ such that
$fE_k$ ($f\sim$ a few$-$ten, in the following discussion, we take
$f= 5$) is the energy injected into the fireball (Zhang et al.
2005). The initial GRB efficiency should be \be
\label{eq:eff_tilde} \tilde{\epsilon}_\gamma \equiv
E_\gamma/(E_\gamma+E_k)= f\epsilon_\gamma/[1+(f-1)\epsilon_\gamma]
\ee where $\epsilon_\gamma\equiv E_\gamma/(E_\gamma+f E_k)$ is the
GRB efficiency derived at $t_d\sim 0.4$. LZ04 find that
$\epsilon_\gamma>0.4$, and therefore, $\tilde{\epsilon}_\gamma>
0.8$, which is too high within the framework of the standard
(internal-shocks) fireball model. However, as shown in \S{2.1},
$\epsilon_\gamma$ presented in LZ04 has been overestimated
significantly. We suggest that $\epsilon_\gamma\sim 0.1$,
therefore even when correcting for the additional energy
$\tilde{\epsilon}_\gamma\sim 0.3$, which is still consistent with
this model.

As an example we consider the $\gamma$-ray efficiency of GRB
050319. Both the optical (Mason et al. 2005) and the X-ray
(Cusumano et al. 2005) light curves are well recorded for this
burst and can be used to constrain the efficiency (see \S
\ref{sec:319} for a detailed discussion).  (1) The time averaged
optical-to-X-ray spectrum ($t\sim 200-900$ s) is a single power
law with an index $\beta=-0.8$ (Mason et al. 2005). This implies
that $\nu_m(t\sim 100~{\rm s})<\nu_R=4.3\times 10^{14}$ Hz. (2)
The very early R-band observation suggests that $F_{\rm
\nu,max}(t\sim 100~{\rm s})\sim 1$ mJy (assuming that energy
injection takes place at $t \ge  400$ s). (3) $\nu_c>\nu_X \sim
10^{17}$ Hz holds up to $t\sim 10^6$ s, as suggested by the XRT
spectrum. We have (see equations
\ref{eq:const3}$-$\ref{eq:const5}) $\epsilon_e\sim 4\times
10^{-2}$, $\epsilon_B\sim 4\times 10^{-5}$, and $E_k\sim 1.3\times
10^{54}$ ergs (the energy carried by the initial outflow). With
K-correction, the isotropic energy of the $\gamma-$ray emission of
GRB 050319 is $E_\gamma \sim 1.2\times 10^{53}$ ergs (Nousek et
al. 2005), so $\tilde{\epsilon}_\gamma=E_\gamma/(E_\gamma+E_k)
\sim 0.1$. It is sufficiently low to be well understood within the
standard fireball model.


\section{Models for a slowly decaying X-ray afterglow}
\label{sec:XRF2}

We turn now to explore (both analytically and numerically) models
that can give rise to a slowly decaying X-ray afterglow phase. The
models we discuss include: (i) Energy injection. (ii) A small
$\zeta_e$. (iii) Evolving shock parameters. (iv)  A very low
nonconstant circum-burst density. We also examine
the possibility of the X-ray
flattening is attributed to a highly magnetized outflow. In the
numerical calculations that we present the parameters are chosen
to reproduce the XRT light curve of GRB 050319 (for $t>$ 380 s).
We also present the corresponding R-band light curve.

\subsection{Energy injection }\label{sec:XRF21}
In the standard fireball model, the fireball that is sweeping the
circum-burst matter decelerates and its  bulk Lorentz factor
evolves with the time as $\Gamma \propto t^{-3/8}$. With
continuous significant energy injection, the fireball decelerates
more slowly and slowly decaying multi-wavelength afterglows are
expected. This model has been analytically investigated by many
authors (Sari \& M\'esz\'aros  2000,  Zhang et al. 2005; Nousek et
al. 2005; Panaitescu et al. 2006; Granot \& Kumar 2006). As shown
in Zhang et al. (2005), for $dE_{inj}/dt \propto t^{-q}$ we find
$\nu_m \propto t^{-(2+q)/2}$, $\nu_c \propto t^{(q-2)/2}$, and
$F_{\rm \nu,max} \propto t^{1-q}$.  In this subsection, we take
$q=0.5$ and find:
\begin{equation} F_\nu \propto
\left\{%
\begin{array}{ll}
    t^{(8-7q)/6}\sim t^{0.75}, & \hbox{for $\nu<\nu_c<\nu_m$;} \\
    t^{(2-3q)/4}\sim t^{1/8}, & \hbox{for $\nu_c<\nu<\nu_m$;} \\
    t^{(8-5q)/6}\sim t^{0.92}, & \hbox{for $\nu_c<\nu<\nu_m$;} \\
    t^{[(6-2p)-(p+3)q]/4}\sim t^{-0.32}, & \hbox{for $\nu_m<\nu<\nu_c$;} \\
    t^{[(4-2p)-(p+2)q]/4}\sim t^{-0.68}, & \hbox{for $\nu>\max \{\nu_c, \nu_m\}$.} \\
\end{array}%
\right.
\end{equation}

\begin{figure}
\begin{picture}(0,200)
\put(0,0){\includegraphics{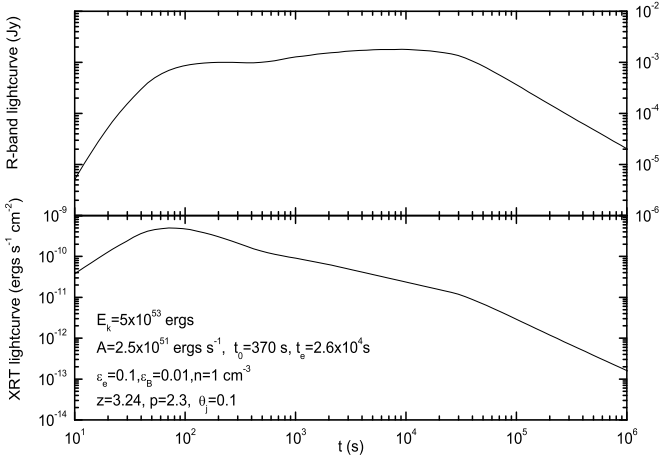}}
\end{picture}
\caption{The X-ray (0.2-10 keV) afterglow light curve and the
R-band light curve for the energy injection model. }
\label{fig:inj1}
\end{figure}

Following Zhang et al. (2005), we consider an  energy injection
rate of the form  $(1+z)dE_{inj}/dt= Ac^2(t/t_0)^{-q}$ for
$t_0<t<t_e$, where $A$ is a constant. With the energy injection,
the equation (8) of Huang et al. (2000) should be replaced by (see
also Wei, Yan \& Fan 2005)
\begin{equation}
d \Gamma={(1-\Gamma^2)dm+{A(t/t_0)^{-q}[dt/(1+z)]}\over
M_{ej}+\epsilon m+2(1-\epsilon)\Gamma m}, \label{Eq:Dyn}
\end{equation}
where $M_{ej}$ is the rest mass of the initial GRB ejecta, $m$ is
the mass of the medium swept by the GRB ejecta, which is governed
by $dm=4\pi R^2 n m_p dR$, $m_p$ is the rest mass of proton,
$dR=\Gamma(\Gamma+\sqrt{\Gamma^2-1})c dt/(1+z)$, $\epsilon=\eta
\epsilon_e$ is the radiation efficiency. Our numerical results,
the R-band emission and the 0.2-10 keV emission, are shown in Fig.
\ref{fig:inj1}.

\subsection{Small $\zeta_e$}\label{sec:XRF22}
In the standard afterglow model, it is assumed that a fraction
$\epsilon_e$ of the shock energy is given to all the fresh
electrons that are swept by the shock front. However, it is
possible that only a fraction $\zeta_e$ of fresh electrons has
been accelerated, as suggested by Papathanassiou \& M\'esz\'aros
(1996). With this correction, equations (\ref{eq:F_max}) and
(\ref{eq:nu_m}) take the form \be F_{\nu,{\rm max}}=6.6~{\rm
mJy}~\zeta_e ({1+z\over 2}) D_{L,28.34}^{-2}
\epsilon_{B,-2}^{1/2}E_{k,53}n_0^{1/2},\label{eq:F_max1} \ee \be
\nu_m = 7.6\times 10^{11}~{\rm Hz}~\zeta_e^{-2} E_{\rm
k,53}^{1/2}\epsilon_{\rm B,-2}^{1/2}\epsilon_{e,-1}^2 C_p^2 ({1+z
\over 2})^{1/2} t_d^{-3/2},\label{eq:nu_m1}\ee respectively.

For $\nu_c<\nu_X<\nu_m$, $F_{\nu_X} \propto t^{-1/4}$. A steeper
decline is possible (the steepest one is $F_{\nu_X} \propto
t^{-4/7}$), depending on the radiative correction, as shown in the
upper panel of Fig. 2 of Sari et al. (1998).

The transition  of the slow decline to a normal decline
($F_{\nu_X}\propto t^{-1.2}$) usually takes place at $t\sim 0.1$
day or earlier, when $\nu_X=\nu_m$. So we have
\begin{equation}
\zeta_e \simeq 0.016 E_{k,53}^{1/4}\epsilon_{e,-1}
\epsilon_{B,-2}^{1/4} t_{d,-1}^{-3/4}C_p[2/(1+z)]^{1/2}.
\end{equation}

\begin{figure}
\begin{picture}(0,200)
\put(0,0){\includegraphics{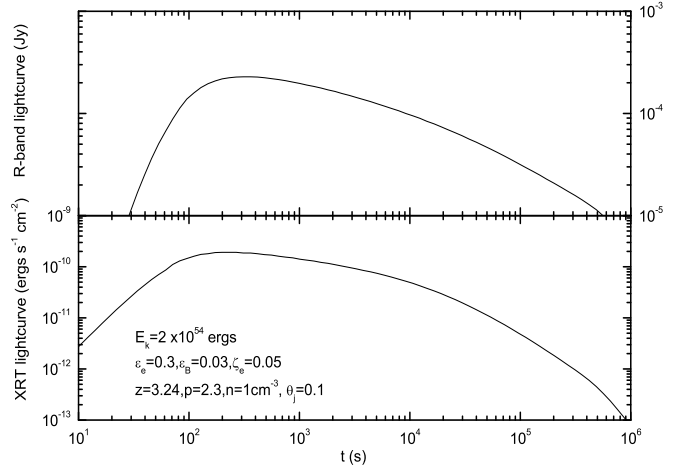}}
\end{picture}
\caption{The  X-ray (0.2-10 keV) afterglow light curve and the
R-band light curve for the small $\zeta_e$ model. }
\label{fig:small-xi}
\end{figure}

The numerical  light curves is  presented in Fig.
\ref{fig:small-xi}. One can see that a long time multi-wavelength
flattening is evident with a small $\zeta_e$.

Before and after the temporal decline transition, the energy
spectrum of the XRT observation should be $F_\nu \propto
\nu^{-1/2}$ and $F_\nu \propto \nu^{-p/2}$, respectively. In other
words, after the break in the light curve, the X-ray spectrum
should be much softer (see also Zhang et al. 2005), which is
inconsistent with most XRT observations (Nousek et al. 2005). In
addition, in this model, the spectral index of the XRT afterglows
in the slow decline phase is $-1/2$. It is much harder than that
of most {\em Swift} X-ray afterglows (see Table 1 of Nousek et al.
2005). {\em The Swift observations therefore provide us robust
evidences of that significant part of, rather than a small
fraction of electrons, have been accelerated in the shock front}.

\subsection{Evolving shock parameters}
\label{sec:XRF23}

In the standard afterglow model, the shock parameters $\epsilon_e$
and $\epsilon_B$ are assumed to be constant. However, it is also
possible that $\epsilon_e$ or $\epsilon_B$, or both, vary with
time  (see Yost et al. (2003) for detailed discussion). Fan et al.
(2002) and Wei et al. (2006) modeled the optical flares detected
in GRB 990123 and GRB 050904 and found  that both $\epsilon_e$ and
$\epsilon_B$ of the forward shock (ultra-relativistic) and reverse
shock (mild-relativistic to relativistic) were very different.
This provides  an indication evidence for a dependence  of the
shock parameters on the strength of the shock. Possible evidence
for the shock strength dependent $\epsilon_B$ was also found by
Zhang, Kobayashi \& M\'esz\'aros (2003), Kumar \& Panaitescu
(2003), McMahon, Kumar \& Panaitescu (2004), Panaitescu \& Kumar
(2004), Fan, Zhang \& Wei (2005b) and Blustin et al. (2006). Yost
et al. (2003) and Ioka et al (2005) considered afterglow emission
assuming $\epsilon_B$ and $\epsilon_e$ are time-dependent,
respectively. Here we simply take $(\epsilon_e,~\epsilon_B)\propto
(\Gamma^{-a},~\Gamma^{-b})~~{\rm for}~\Gamma>\Gamma_o, ~{\rm
otherwise}~~(\epsilon_e, \epsilon_B)\sim {\rm const.}$, where
$\Gamma_o$ is the Lorentz factor of the outflow at the X-ray
decline translation, both $a$ and $b$ are taken to be positive.
For simplicity, we discuss only the case of $a=b$ for
$\Gamma>\Gamma_o$. Below $\Gamma_o$, the solution is the usual
one.

The typical synchrotron radiation frequency $\nu_m$ satisfies
\begin{eqnarray}
\nu_m  \propto (\Gamma/\Gamma_o)^{-5a/2} t^{-3/2} \propto
t^{(15a-24)/16}, \label{eq:nu_m2}
\end{eqnarray}
where $\Gamma \approx 25 E_{\rm
iso,53}^{1/8}[2/(1+z)]^{-3/8}t_{d,-1}^{-3/8}n_0^{-1/3}$.

The cooling frequency $\nu_c$ satisfies
\begin{eqnarray}
\nu_c \propto (\Gamma/\Gamma_o)^{3a/2}t^{-1/2} \propto
t^{-(8+9a)/16}, \label{eq:nu_c2}
\end{eqnarray}
The maximum spectral flux $F_{\nu,{\rm max}}$ satisfies
\begin{eqnarray}
F_{\nu,{\rm max}} \propto (\Gamma/\Gamma_o)^{-a/2} \propto
t^{3a/16}. \label{eq:F_max2}
\end{eqnarray}
The observed flux behaves  as (in this subsection, we take $a=1$):
\begin{equation}
F_{\nu_X} \propto   \left\{%
\begin{array}{ll}{F_{\rm \nu,max}\nu_c^{-1/3}\propto t^{{4+9a\over 24}}}\sim t^{0.55}, &
\\ \qquad \qquad \qquad \qquad \qquad \qquad  \hbox{for $\nu_X<\nu_c<\nu_m$;} \\
    F_{\rm \nu,max}\nu_c^{1/2}\propto t^{-{8+3a \over32}}\sim t^{-0.35}, &
\\ \qquad \qquad\qquad \qquad \qquad \qquad \hbox{for $\nu_c<\nu_X<\nu_m$;} \\
    F_{\rm \nu,max}\nu_m^{\rm -1/3}\propto t^{{4-a \over 8}}\sim t^{0.38}, &
\\ \qquad \qquad \qquad \qquad\qquad \qquad \hbox{for $\nu_X<\nu_m<\nu_c$;} \\
    F_{\rm \nu,max}\nu_m^{{p-1 \over 2}}\propto
t^{{15ap-9a-24p+24\over 32}}\sim t^{-0.2}, &
\\ \qquad \qquad \qquad \qquad \qquad \qquad\hbox{for $\nu_m<\nu_X<\nu_c$;} \\
   F_{\nu,{\rm max}} \nu_c^{1/2}\nu_m^{{p-1\over 2}}\propto t^{{16-18a-24p+15ap \over 32}}
   \sim t^{-0.65}, &
\\ \qquad \qquad \qquad \qquad  \qquad \qquad \hbox{for $\nu_X>\max\{ \nu_c, \nu_m \}$.} \\
\end{array}%
\right.
\end{equation}

\begin{figure}
\begin{picture}(0,200)
\put(0,0){\includegraphics{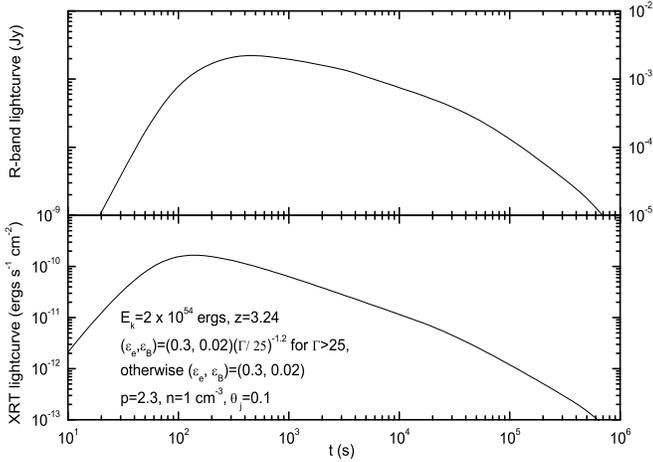}}
\end{picture}
\caption{The X-ray (0.2-10 keV) and R-band afterglow light curves
for the evolving shock parameter model. The parameters are listed
in the figure.} \label{fig:shock-parameter}
\end{figure}

The afterglow light curves are shown in Fig.
\ref{fig:shock-parameter}.  As  both $\epsilon_e$ and $\epsilon_B$
increase with time, the flux of the early X-ray emission is dimmer
than that of the constant shock parameters model and the decline
is much slower. Both are consistent with the current {\em Swift}
XRT observations (Nousek et al. 2005).

\subsection{A very low nonconstant density}
\label{sec:XRF25} In the standard ISM afterglow model, the number
density of the medium is taken as a constant. In the wind model,
the number density $n$ decreases with the radius $R$ as $n\propto
R^{-2}$ (M\'esz\'aros, Rees \& Wijers 1998; Dai \& Lu 1998;
Chevalier \& Li 2000). Here we discuss the general case $n\propto
R^{-k}~(0 \leq k<3)$.

First, we show that for a fireball decelerating in the BM
self-similar regime (Blandford \& McKee 1976),  no X-ray
flattening is expected regardless of the choice of $k$. The energy
of the fireball is nearly  constant and it is given by $E_{\rm
iso}\approx \Gamma^2 M c^2$, where $M\propto R^{3-k}$ is the total
mass of the swept medium. So $\Gamma \propto R^{-(3-k)/2}$.
Considering that $dR \propto \Gamma^2 dt \propto R^{-(3-k)}dt$, we
have $R \propto t^{1/(4-k)}$ and $\Gamma \propto
t^{-(3-k)/[2(4-k)]}$.

Now $\nu_m$ decreases with $t$ as
\begin{equation}
\nu_m \propto \Gamma^4 R^{-k/2}\propto t^{-3/2},
\end{equation}
and $\nu_c$ and $F_{\nu,{\rm max}}$ satisfy
\begin{equation}
\nu_c \propto \Gamma^{-4}R^{3k/2}t^{-2}\propto
t^{(3k-4)/[2(4-k)]},
\end{equation}
\begin{equation}
F_{\nu,{\rm max}}\propto R^{3-3k/2}\Gamma^2 \propto
t^{-k/[2(4-k)]},
\end{equation}
respectively. This results in:
\begin{equation}
F_{\nu_X}\propto \left\{%
\begin{array}{ll}
    t^{-3(p-1)/4-k/[2(4-k)]}, & \hbox{for $\nu_m<\nu_X<\nu_c$;} \\
    t^{-1/4}, & \hbox{for $\nu_c<\nu_X<\nu_m$;} \\
    t^{-(3p-2)/4}, & \hbox{for $\nu_X>\max \{ \nu_m, \nu_c \}$.} \\
\end{array}%
\right.
\end{equation}
The last two are independent of $k$. So no X-ray flattening
appears.

\begin{figure}
\begin{picture}(0,200)
\put(0,0){\includegraphics{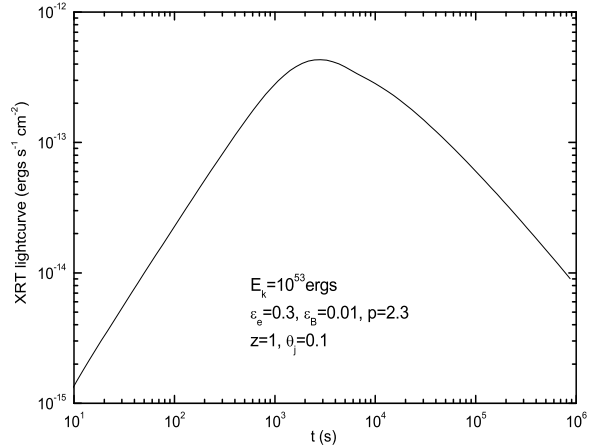}}
\end{picture}
\caption{X-ray (0.2-10 keV) afterglow light curve for a very low
nonconstant density:  $n=10^{-4}~{\rm cm^{-3}}$ for $R<10^{16}$
cm; $n=10^{-4}~(R/10^{16})^{-1}~{\rm cm^{-3}}$ for
$10^{16}<R<10^{19}~{\rm cm}$ and $n=10^{-7}~{\rm cm^{-3}}$  for
$R>10^{19}$ cm.} \label{fig:density}
\end{figure}

However, if the number density is sufficiently low, the
deceleration timescale ($\propto n^{-1/3}$) can be very long and
even as long as $\sim 10^4$ s. In this case, a slowly decaying
X-ray afterglow may be obtained. One example has been plotted in
Fig. \ref{fig:density}, in which the density profile of the medium
is taken as $n=10^{-4}~{\rm cm^{-3}}$ for $R\leq 10^{16}$ cm,
$n=10^{-4}R_{16}^{-1}~{\rm cm^{-3}}$ for $1 \leq R_{16} \leq
10^3$, and $n=10^{-7}~{\rm cm^{-3}}$ for $R_{16}>10^3$. An X-ray
flattening appears when the shock front reaches $R=10^{19}$ cm.
However, while the shape of the light curve is correct the X-ray
flux is too low to account for most XRT light curves.

\subsection{Magnetized outflow }\label{sec:XRF26}
A Poynting flux dominated outflow  (Usov 1994; Thompson 1994;
Lyutikov \& Blandford 2003) is an alternative to the standard
baryonic fireball model. Within the context of this discussion it
is of interest since it may also give rise to a slowly decaying
X-ray afterglow (Zhang et al. 2005). We investigate, here, briefly
this possibility, extended discussion will be presented elsewhere.

We assume that the electromagnetic energy $E_p$  will be
transformed continuously into the kinetic energy of the forward
shock.  The dynamical evolution of the shocked medium is governed
by (Huang et al. 2000; Wei et al. 2006):
\begin{equation}
d\Gamma=-{(\Gamma^2-1)dm+dE_{p}/c^2 \over M_{\rm ej}+\epsilon m
+2(1-\epsilon)\Gamma m}, \label{eq:mag1}
\end{equation}
where $E_p\equiv \Gamma^2 VB'^2/(4\pi)$, $V$ is the volume of the
magnetized outflow (measured by the observer) and $B'$ is the
comoving strength of the magnetic field.

If the magnetic pressure is higher than the thermal pressure of
the shocked medium, the magnetic pressure works on the shocked
medium and the kinetic energy of the forward shock increases. A
 pressure balance between the shocked medium and the magnetized
 outflow is established, so we
 have (see also Lyutikov \& Blandford 2003)
 $B'^2/(8\pi)=p_{gas}\simeq 4\Gamma^2 n m_p c^2/3$,
where $p_{gas}$ is the thermal pressure of the shocked medium.
Therefore $E_p$ can be estimated by\footnote{Providing that
$V\propto R^{\rm 2+c}\Gamma^{\rm -d}$, $E_p \propto \Gamma^{\rm
8+2c-d}t^{\rm 2+c} \propto t^{-\delta}$ ($c$, $d$, and $\delta$
are all larger than 0), we have $\Gamma \propto t^{\rm
-(2+c+\delta)/(8+2c-d)}$, which should be flatter than
$\Gamma^{-3/8}$ (the canonical dynamical evolution of a ejecta
without energy injection). It requires that $2c+3d<8-8\delta$,
 otherwise $dE_p/dt<0$ has been violated. It is evident that in the
 spreading phase, i.e., $c=1$ and $d=2$, $E_p$ can not be converted into
 the kinetic energy of the forward shock effectively.}
\begin{equation}
E_p=2\Gamma^2 P_{gas}V\approx 8\Gamma^4 n m_p c^2 V/3.
\label{eq:mag2}
\end{equation}

$dE_p/c^2$ can be calculated as follows. Assuming that the whole
system (the shocked medium and the magnetized outflow) is
adiabatic (i.e., the radiation efficiency $\epsilon=0$), the
energy conservation yields \be 2\Gamma^2 p_{gas}V+3\Gamma^2
p_{gas} (V_{tot}-V)=E_{tot}-\Gamma (M_{\rm ej}+m)
c^2,\label{eq:mag3} \ee where $V_{tot}\approx 4\pi R^2 \Delta$ is
the total volume of the system, $\Delta$ is the width of the
system, which is described by $d \Delta=(\beta_{fsh}-\beta)dR$ and
$\beta_{\rm fsh}\simeq \sqrt{\Gamma^2-1}/[\Gamma-{1\over
4\Gamma+3}]$ is the velocity of the forward shock. Differentiating
equation (\ref{eq:mag3}) we obtain:
\begin{eqnarray}
dE_p/c^2 &=& 2\{(16 \Gamma^3 n m_p V_{tot}+M_{\rm ej}+m)d\Gamma +
16\pi \Gamma^4 R n m_p \nonumber\\
&& [R(\beta_{fsh}-\beta) +2\Delta] dR+\Gamma dm\}.
\end{eqnarray}

After simple algebra, equation (\ref{eq:mag1}) can be rearranged
as: (note that now we take $\epsilon=0$)
\begin{eqnarray}
d\Gamma=-{(\Gamma^2+2A\Gamma-1)dm
  +B dR \over M_{\rm ej}+2\Gamma m+ 2A (16\Gamma^3 n m_p
V_{tot}+M_{\rm ej}+m)}, \label{eq:mag4}
\end{eqnarray}
where $A=1$ for $dE_p/dR \leq 0$, otherwise $A=0$; $B=32A\pi n m_p
\Gamma^4 R [(\beta_{fsh}-\beta)R+2\Delta]$.

With proper boundary conditions and the relations $dm=4\pi n m_p
R^2dR$, $dR=\beta c \Gamma^2(1+\beta)dt/(1+z)$, $V_{tot}=4\pi
R^2\Delta$ and $d\Delta=(\beta_{fsh}-\beta)dR$, equation
 (\ref{eq:mag4}) can be solved numerically. In our numerical example, we take
 $E_k=10^{52}$ ergs, $n=1~{\rm cm^{-3}}$, $E_p=10E_k$ and the width of the
 outflow is taken as $3\times 10^{11}$ cm. The starting point of
 our calculation is at $R=2\times 10^{16}$ cm ($\sim R_{dec}$, the
 deceleration radius of the outflow, where $\sim E_k/2$ has been
 given to the shocked medium), at which
 $\Gamma=360$ \footnote{At that radius, the reverse shock has crossed the
 ejecta and a pressure balance between the shocked medium and the
 magnetized outflow is reached. In this work, we do not calculate
 the reverse shock emission (see Fan, Wei \& Wang (2004a)
 for the reverse shock emission with mild magnetization and Zhang \& Kobayashi (2005)
 for reverse shock emission with arbitrary magnetization). With the
 ideal MHD jump condition, the reverse shock can not convert the magnetic
 energy into the kinetic energy of the forward shock effectively, as shown in
 Kennel \& Coronitti (1984), Fan, Wei \& Zhang (2004b) and
 Zhang \& Kobayashi (2005) both analytically and numerically.}.
 We find out that
 most of the magnetic energy has been converted into the kinetic energy
 of the forward shock in a very short time $\sim 50(1+z)$ s. A similar result has been obtained by
 Lyutikov \& Blandford (2003). Though this timescale is much longer than the crossing time
 of the reverse shock, it is not long enough to account for the X-ray flattening detected in most GRBs.


\section{Case studies: constraining the models}\label{sec:0319}
GRB 050319 and GRB 050401, have well recorded X-ray and optical
afterglows, with which the models discussed in \S\ref{sec:XRF2}
can be constrained. We discuss these constraints in detail here.
For most {\em Swift} GRBs only the X-ray afterglow is well
detected. Such bursts provide, of course, much weaker constraints
on the model . We discuss one example, GRB 050315, briefly.

\subsection{GRB 050319}\label{sec:319}
Both the optical and X-ray afterglows of GRB 050319 have been well
recorded (Wo\'zniak et al. 2005; Mason et al. 2005; Cusumano et
al. 2005; Nousek et al. 2005). The optical flux declines with a
power law slope of $\alpha=-0.57$ between $\sim 200$s after the
burst onset until it fades below the sensitivity threshold of the
UVOT after $5\times 10^4$ s. The optical V-band emission lies on
the extension of the X-ray spectrum, with an spectral slope
$\beta=-0.8$ (Mason et al. 2005). The temporal behavior of the
X-ray afterglow is more complicated. After a steep decay ($\alpha=
-5.53$) up to $t=370$s, the light curve shows a slow decay with a
temporal index of $\alpha=-0.54$. It steepens to $\alpha=-1.14$ at
$t=2.60\times 10^4$s. The spectral indices in the slow decline
phase and the normal decay phase are $\beta=-0.7$ and
$\beta=-0.8$, respectively (Cusumano et al. 2005; Nousek et al.
2005; However, see Quimby et al. 2006). Below we examine whether
the models discussed above (in \S\ref{sec:XRF2}) can explain both
the optical and the X-ray afterglows self-consistently.

{\bf Energy injection:} The energy injection model is believed to
able to explain the observation  (e.g., Zhang et al. 2005; Mason
et al. 2005; Cusumano et al. 2005).  As shown in
\S\ref{sec:XRF21}, for $q=0.6$ and $p=2.4$, both the optical and
the X-ray afterglows decline as $F_{\nu_X}\propto t^{-0.54}$ when
$\nu_m<\nu_R<\nu_X<\nu_c$, the corresponding spectral index should
be $\beta=-(p-1)/2\sim -0.7$. All these values are consistent with
the observation. However, the non-detection of the further X-ray
break caused by the spectral translation ($\nu_c<\nu_X$) up to
$\sim 10^6$ s after the trigger suggests that $\epsilon_B\sim
5\times 10^{-3}$ and $n\sim 10^{-3}~{\rm cm^{-3}}$ (Cusumano et
al. 2005). The problem is that $F_{\nu_X}$ depends on $n$ and
$\epsilon_B$ sensitively for $\nu_X<\nu_c$ (see equation
(\ref{eq:F2})). The smaller $n$ and $\epsilon_B$, the smaller
$F_{\nu_X}$. We show below that it is quite difficult to reproduce
the detected X-ray and optical light curves with the energy
injection model.

The earliest R-band data is collected at $\sim 200$ s (note that
the real onset of GRB 050319 is about 130 s before the {\em Swift}
trigger time taken in Wo\'znik et al. (2005), see Cusumano et al.
(2005) for clarification), and the flux is about $F_{\nu_R}\sim
0.7$ mJy. At that time, the total energy of the outflow is still
dominated by the initial $E_k$, and $F_{\rm \nu,max}$ and $\nu_m$
are still described by equations (\ref{eq:F_max}) and
(\ref{eq:nu_m}), respectively. The conditions $F_{\rm \nu,max}\geq
0.7$ mJy and $\nu_m (t\sim 200 ~{\rm s})\leq \nu_R$ yield
\begin{eqnarray}
\epsilon_{B,-2}^{1/2}E_{k,53}n_0^{1/2} \geq   0.8 &\Rightarrow &
E_{k,53}\geq 0.8\epsilon_{B,-2}^{-1/2}n_0^{-1/2},\label{eq:const1} \\
E_{k,53}^{1/2}\epsilon_{B,-2}^{1/2}\epsilon_{e,-1}^2 \leq 0.04
&\Rightarrow & \epsilon_e\leq 0.02
E_{k,53}^{-1/4}\epsilon_{B,-2}^{-1/4}\label{eq:const2},
\end{eqnarray}
respectively. The condition $\nu_c \geq \nu_X \sim 10^{17}$ Hz
holding up to $t \sim 10^6$ s gives
\begin{eqnarray}
&& E_{k,53}^{-1/2}\epsilon_{B,-2}^{-3/2}n_0^{-1} (1+Y_o)^{-2}\geq
940\nonumber\\ && \Rightarrow \epsilon_B\leq 10^{-4}
E_{k,53}^{-1/3}n_0^{-2/3}(1+Y_o)^{-4/3},\label{eq:const3}
\end{eqnarray}
where $Y_o$ is the Compton parameter at $t\sim 10^6$ s.  To
derived this relation, we assume that at $t\sim 200$ s, $\nu_c$ is
described by equation (\ref{eq:nu_c}), and $\nu_c \propto
t^{(q-2)/2}\sim t^{-0.7}$ up to $t\sim 2.6\times 10^4$ s (i.e., in
the energy injection phase), then $\nu_c\propto t^{-1/2}$ up to
$t\sim 10^6$ s. Combing equations
(\ref{eq:const1}-\ref{eq:const3}), we have
\begin{eqnarray}
E_{k,53}&\geq & 13 n_0^{-1/5}(1+Y_o)^{4/5}, \label{eq:const4}\\
\epsilon_e &\leq& 0.06
E_{k,53}^{-1/6}n_0^{1/6}(1+Y_o)^{1/3}\label{eq:const5}.
\end{eqnarray}
 Now $Y_0 \sim 0$ (see the Appendix for detail), we have
$E_k>1.3\times 10^{54}~{\rm ergs}~n_0^{-1/5}$. On the other hand,
the energy injection coefficient $A'\equiv Ac^2 \sim (1+z)E_k/t_0
\sim 1.4\times 10^{52}n_0^{-1/5}~{\rm ergs~s^{-1}}~$  for $t_0
\sim 370$ s. Please note that $A'$ is comparable to the recorded
luminosity of most GRBs and the X-ray luminosity recorded by XRT
is just $\sim 10^{48}~{\rm ergs~s^{-1}}$. The outflow accounting
for the late time injection is so energetic that strong soft X-ray
to $\gamma-$ray emission powering by shocks or magnetic
dissipation are expected. They will quite likely dominate over the
corresponding forward shock emission, which is inconsistent with
the observation.

This model is also disfavored by the different temporal behavior
of the X-ray and the optical afterglows at $t>2.6\times 10^{4}$ s.
We therefore conclude that the energy injection model can't
account for the multi-wavelength afterglows of GRB 050319. We
tried to fit both the R-band and X-ray afterglows with reasonable
parameters numerically but failed.

Provided that the energy injection model works (i.e., there is a
mechanism to keep such energetic outflow steady enough and there
is no magnetic dissipation), the initial GRB efficiency in this
case is as low as $\tilde{\epsilon}_\gamma=
E_\gamma/(E_\gamma+E_k) \sim 0.08~n_0^{1/5}$.

{\bf Small $\zeta_e$:} This model is disfavored by two facts. One
is that in the X-ray flattening phase, $\nu_c<\nu_X<\nu_m$, the
corresponding spectral index  is $\beta \sim -1/2$, which only
marginally matches the observation $\sim -0.7$. The other is that
after the temporal transition at $t\sim 10^4$ s,
$\nu_c<\nu_m<\nu_X$, the spectral index should be $\beta=-p/2 \sim
-1.2$, which is inconsistent with the observation.

{\bf Evolving shock parameters:} As shown in \S{\ref{sec:XRF23},
for $\nu_m<\nu_R<\nu_X<\nu_c$, $p=2.4$ and $a=b=0.6$,
$(F_{\nu_R},~F_{\nu_X})\propto t^{-0.54}$ and the spectral index
$\beta=-(p-1)/2\sim -0.7$, are all consistent with the
observation. After the shock parameters saturate at $t\sim
2.6\times 10^4$ s, $F_{\nu_X}\propto t^{-1.1}$ and $\beta=-0.7$ as
long as $\nu_X<\nu_c$, which also matches the observation.
However, the optical light curve should be much steeper since
$\nu_m<\nu_R<\nu_c$ also holds. The UVOT observation and the
ground based R-band observation suggest that the decline of
optical emission does not change up to $t\sim 2\times 10^5$ s,
though the scatter of the flux is quite large (Kiziloglu et al.
2005; Sharapov et al. 2005; see Mason et al. 2005 for a summary).
Therefore the evolving shock parameter model is disfavored.

{\bf Very low nonconstant density:} With proper parameters as well
as proper density profile, an X-ray flattening does appear (see
Fig.\ref{fig:density}). However, as already mentioned, the flux is
too low to match most observations, here we do not discuss it
further.

{\bf Off-beam annular jet model.} Recently, Eichler \& Granot
(2005) suggested that the flat part of the XRT light curve may be
a combination of the decaying tail of the prompt $\gamma-$ray
emission and the delayed onset of the afterglow emission observed
from viewing angles slightly outside of the edge of the jet (i.e.,
off-beam). This model, like others mentioned above, can account
for the slow decline of many X-ray afterglows, but may be unable
to explain both the optical and the X-ray afterglows of GRB 050319
self-consistently, as shown below.

Following Eichler \& Granot (2005), we assume that the off-beam
angle is $\delta \theta \sim 1/\Gamma_{int}$, where $\Gamma_{int}$
is the initial Lorentz factor of the outflow. Larger $\delta
\theta$ is less favored since the slowly decayingR-band afterglow
has been well recorded as early as $t\sim 200$ s, which implies
that the afterglow onset has not been delayed so much. In the
off-beam case, the typical synchrotron radiation frequency should
be
\begin{equation} \nu_m \approx  7.6\times 10^{11}~{\rm Hz}~E_{\rm
k,53}^{1/2}\epsilon_{\rm B,-2}^{1/2}\epsilon_{e,-1}^2 C_p^2 ({1+z
\over 2})^{1/2} a^{1/2} t_d^{-3/2},
\end{equation}
where $a\approx [1+(\Gamma_{int}\delta \theta)^2] \sim 2$ is the
Doppler factor. Therefore the condition  $\nu_m (t\sim 200~{\rm
s})<\nu_R$ results in
\begin{equation}
\epsilon_e\leq 0.017
E_{k,53}^{-1/4}\epsilon_{B,-2}^{-1/4}(a/2)^{-1/4}. \label{eq:EG1}
\end{equation}
For $\delta \theta\sim 1/\Gamma_{int}$, the late time (i.e., the
normal decline phase) afterglow emission is quite similar to the
on-beam case (Eichler \& Granot 2005).

We use equation (\ref{eq:F2}) to estimate the late time X-ray
flux, though the predicted flux of the annular jet model should be
somewhat different from that of our conical jet model (Granot
2005; Eichler \& Granot 2005). The XRT flux $\approx 8\times
10^{-12}~{\rm ergs~s^{-1}~cm^{-2}}$ at $t_d\sim 0.3$ gives
\begin{equation}
E_{k,53}\approx
0.33\epsilon_{e,-1}^{4(1-p)/(p+3)}\epsilon_{B,-2}^{-(p+1)/(p+3)}n_0^{-2/(p+3)}.\label{eq:EG2}
\end{equation}
The condition $\nu_c>\nu_X\sim 10^{17}$ Hz holding up to $t\sim
10^6$ s yields
\begin{equation}
\epsilon_B <3\times 10^{-4}
E_{k,53}^{-1/3}n_0^{-2/3}.\label{eq:EG3}
\end{equation}
Combing equations (\ref{eq:EG1}-\ref{eq:EG3}), we have $E_k >
10^{55}~{\rm ergs}~n_0^{-1/5}(a/2)^{3(p-1)/10}$. While we manage
to fit both X-ray and optical data, the energy needed is too large
for any realistic progenitor models. We therefore suggest that the
off-beam annular jet model is also unable to account for the
afterglows of GRB 050319.

\subsection{GRB 050401}
The early X-ray light curve is consistent with a broken power law
with $\alpha=-0.63$ and $-1.41$ respectively, the break is at
$t_b\sim 4480$ s (de Pasquale et al. 2006). The X-ray spectral
indices before and after the break are nearly constant $\sim
-0.90$. Therefore the small $\zeta_e$ model is ruled out directly.
Zhang et al. (2005) also show that the flat electron distribution
model ($1<p<2$) is unable to account for the X-ray afterglow
observation. The afterglow has also been detected in R-band, which
decays as a simple power law $\propto t^{-0.76}$ up to $t\sim
3.5\times 10^4$ s (Rykoff et al. 2005).

{\bf Energy injection ($p\sim 2.8$).}  After the break, the light
curve is consistent with an ISM model for $\nu_m < \nu_X <\nu_c$
with $p \sim 2.8$. Before the break, it is consistent with the
same model with $q=0.5$ (see also Zhang et al. 2005). As far as
the R-band afterglow emission is concerned, there are two
possibilities. One is that $\nu_m<\nu_R<\nu_c$, the optical
afterglow should follow the temporal behavior of the X-ray
afterglow, which is not the case. The other is that $\nu_R<\nu_m$
for $t \leq t_b$, the afterglow increases as $t^{0.9}$ for $q\sim
0.5$ (see \S\ref{sec:XRF21}), which is inconsistent with the
observation. We therefore conclude that the popular energy
injection model is unable to account for the data in this burst as
well.

{\bf Evolving shock parameters  ($p\sim 2.8$)}. The light curve
after the break is consistent with an ISM model for $\nu_m < \nu_X
<\nu_c$ with $p \sim 2.8$. Before the break, it is consistent with
the same model with $a=b=0.7$. Can it reproduce the optical
afterglow? The answer is negative. Provided that $\nu_R<\nu_m$ for
$t \leq t_b$, the optical afterglow should increase as $t^{0.4}$
(see \S\ref{sec:XRF23}), which is inconsistent with the data. The
case of $\nu_m<\nu_R$ is ruled out directly in view of the
different temporal behavior of X-ray and R-band afterglows.

\subsection{GRB 050315}
After a steep decay up to $t_{b1}=308$ s, the X-ray light curve
shows a flat ``plateau'' with a temporal index of $\alpha=-0.06$
(the spectral index  of XRT data is $\beta=-0.73$). It then turns
to $\alpha=-0.71$ at $t_{b2}=1.2\times 10^4$ s, the spectral index
is $\beta=-0.79$. Finally there is a third break at
$t_{b3}=2.5\times 10^5$ s, after which the temporal decay index is
$\alpha=-2.0$ and the spectral index is $\beta=-0.7$ (Nousek et
al. 2005; Barthelmy et al. 2005).

There are two possible interpretations for the long term constant
spectral index $\beta \sim -0.7$. One is that
$\max\{\nu_c,\nu_m\}<\nu_X$ after $t=308$ s and the power law
index of the shocked electron $p\sim 1.5$. The other is that
$\nu_m<\nu_X<\nu_c$ for $t_{b1}<t<t_{b3}$ and $p\sim 2.5$.

 {\bf Energy injection  ($p\sim 2.5$)}. To obtain the
 slow decline for $t_{b1}<t<t_{b2}$, energy injection with $q\sim 0.2$ is
 needed. $q\sim 0.9$ is needed to
reproduce the X-ray afterglows at $t_{b2}<t<t_{b3}$. The late time
sharp decay appears when the boundary of a
 non-spreading jet becomes visible.

 {\bf Evolving shock parameters  ($p\sim 2.5$)}. As shown in
 \S\ref{sec:XRF23}, with $a=b=1.2$, we have a slow decline
 slope $\alpha=-0.06$ between $t_{b1}$ and $t_{b2}$. To get a decline
 slope $\alpha=-0.71$ between $t_{b2}$ and $t_{b3}$, $a=b=0.45$ are
 needed. The late time sharp decay appears when the boundary of a
 non-spreading jet becomes visible and $a=b=0.45$.

 We find that both models can explain the observed X-ray light
 curves of GRB 050315.


\section{Summary \& Discussion}
\label{sec:XRF4}

During the past several months, the {\it Swift} XRT has collected
a rich sample of early X-ray afterglow data. A good fraction of
these afterglows show a slow decline phase lasting between a few
hundred to several thousand seconds. The energy injection model is
the leading model to account for these slowly decaying X-ray
afterglows (e.g., Zhang et al. 2005; Nousek et al. 2005;
Panaitescu et al. 2006; Granot \& Kumar 2006). It has been
suggested that in this model, the GRB efficiency might be as high
as 90\%.  Such a high efficiency challenges the standard internal
shock model for the prompt $\gamma-$ray emission.

In this work, we have re-examined  the GRB efficiency of several
pre-{\it Swift} GRBs and one {\it Swift} GRB. In addition, we have
explored several mechanism which might give rise to a slowly
decaying X-ray light curve and we have compared the predictions of
these models with the well recorded multi-wavelength afterglows of
GRB 050319 and GRB 050401. We  draw the following conclusions:

1. The GRB efficiency of pre-{\it Swift} GRBs that has been
derived directly from the X-ray flux 10 hours after the burst has
been overestimated. For these {\it Swift} GRBs with long time
X-ray flattening, the GRB efficiency is also moderate (around
0.5), even when taking into account the possibility of energy
injection. Such efficiency can be understood within the standard
internal shock model.

2. With a proper choice of parameters, the slow decline slope of
X-ray afterglow like the one  detected in GRB 050319 can be well
reproduced by several models---the energy injection model,
evolving shock parameter model (in which the shock parameters are
assumed to increase with the decrease of the shock strength for
$t<10^4$ s), the small $\zeta_e$ model (in which the shock energy
has been give to a fraction $\zeta_e$ of electrons, rather than
total) and the very low nonconstant density model. Out of these
models, the last two are ruled out by the X-ray data itself. In
the last model, the resulting X-ray afterglow is too dim to match
most XRT observations. The small $\zeta_e$ model is also
disfavored since (1) In the slow decline phase, the XRT spectrum
are usually much softer than $\nu^{-1/2}$; (2) After the light
curve break, no spectral steepening has been detected in most
cases, which is inconsistent with the model. The other models,
including the energy injection model and the evolving shock
parameter model seem to be consistent with the X-ray afterglow
observations.

3. While two  models: the energy injection model and  the evolving
shock parameter model are consistent with the X-ray data, they
fail to reproduce both the X-ray and the optical afterglows of GRB
050319 and GRB 050401. In each burst, the optical flux declines
slowly up to $\sim 10^5$ s. On the other hand, the X-ray light
curve decays slowly up to $t\sim 10^4$ and then turns to the
normal faster decay ($F \propto t^{-1.2}$ or so). The temporal
index of the slow decay X-ray phase is close to that of the
optical light curve. The XRT spectrum is unchanged before and
after the X-ray break. This means that the break is not caused by
a cooling break in which $\nu_c$ crosses the observed frequency. .

The failure of all models that we considered to fit both the X-ray
and the optical afterglow light curves suggests that we should
look for another alternative. An intriguing possibility is based
on fact that the extrapolation backwards of the late X-ray light
curve is in agreement with (or 1 to 2 order lower than) the prompt
X-ray emission. This suggests that we face a "missing energy
problem". Namely, during the slow decay phase (in which the X-ray
flux is rather low) we miss X-ray emission. Is it possible that
during this phase this energy is dissipated into a different
channel and not into Synchrotron X-rays and that this different
channel becomes ineffective at around ten hours? Put differently,
during this phase the electrons within the forward shock emit
Synchrotron X-rays inefficiently. A possibility of this kind (that
we have considered and found not to work) is if the X-ray emitting
electrons are cooled efficiently via inverse Compton (and hence
their Synchrotron X-ray emission is weaker). As already mentioned
inverse Compton cooling is important in determining the X-ray
flux. Furthermore, due to the Klein-Nishina cutoff this cooling
becomes unimportant at approximately one day. However, this
transition is not sharp enough to produce the observed slowly
decaying X-ray light curves. While inverse Compton cooling does
not work it is possible that another, yet unexplored, process of
this kind is responsible for the observed light curves.


\section*{Acknowledgments}
We thank A. Panaitescu and E. Waxman for their constructive
comments. We also thank the referee for helpful suggestion. Y. Z.
Fan thanks D. M. Wei for his help and J. Dyks for checking Fig.
\ref{fig:density}. This work is supported by US-Israel BSF. TP
acknowledges the support of Schwartzmann University Chair. YZF is
also supported by the National Natural Science Foundation (grants
10225314 and 10233010) of China, and the National 973 Project on
Fundamental Researches of China (NKBRSF G19990754).

\begin{appendix}
\section{The general form of the inverse Compton
parameter}\label{sec:Y} For the photons with frequency higher than
$\hat{\nu}$, the Compton parameter should be suppressed
significantly since it is the Klein-Nishina regime, where
$\hat{\nu}$ is governed by $(1+z)\gamma_e h \hat{\nu} \sim \Gamma
m_e c^2$, i.e.,
 \begin{equation}
 \hat{\nu}\sim 1.2\times 10^{20}~{\rm
Hz}~(1+z)^{-1}\Gamma \gamma_e^{-1}.
\end{equation}

We extend the derivation of the Compton parameter $Y$ given by
Sari et al. (1996) to the general form, in the limit of single
scattering. The ratio of the inverse Compton power ($P_{\rm IC}$)
to the synchrotron power ($P_{\rm syn}$) of an electron with
random Lorentz factor $\gamma_e$ is given by
\begin{equation}
Y(\gamma_e)={P_{\rm IC}\over P_{\rm syn}}={\eta_{_{KN}} U_{\rm
syn} \over U_{\rm B}}={\eta \eta_{_{KN}}\epsilon_e \over
[1+Y(\gamma_e)]\epsilon_B},
\end{equation}
where $\eta_{_{KN}}$ is the fraction of synchrotron radiation
energy of total electrons emitted at frequencies below
$\hat{\nu}$. So we have
\begin{equation}
Y(\gamma_e)=(-1+\sqrt{1+4\eta \eta_{_{KN}}
\epsilon_e/\epsilon_B})/2.
\end{equation}
Below we estimate the parameter $\eta_{_{KN}}$ in different
cooling regimes.

{\bf A. Slow cooling.}
\begin{equation} F_\nu =F_0
\left\{%
\begin{array}{ll}
    (\nu/\nu_c)^{-(p-1)/2}, & \hbox{for $\nu_m<\nu<\nu_c$;} \\
    (\nu/\nu_c)^{-p/2}, & \hbox{for $\nu_c<\nu<\nu_M$.} \\
\end{array}%
\right.
\end{equation}
where $\nu_M\sim 2.8\times 10^{22}\Gamma /(1+z)~{\rm Hz}$ is the
maximal synchrotron radiation frequency of the electrons
accelerated by the forward shock. For $p>2$, the total energy
emitted is $\int F_\nu d\nu = {2F_0\over (3-p)}\nu_c^{(p-1)/2}
[{1\over (p-2)} \nu_c^{(3-p)/2} - \nu_m^{(3-p)/2}]$, where the
photons with frequencies below $\nu_m$ have been ignored.
Throughout the Appendix, $\nu_c$ is still described by equation
(\ref{eq:nu_c}) but without the correction of $1/(1+Y)^2$. We have
\begin{equation} \eta_{_{KN}}\sim
\left\{%
\begin{array}{ll}
    0, & \hbox{for $\hat{\nu}<\nu_m$;} \\
    {\hat{\nu}^{(3-p)/2}-\nu_m^{(3-p)/2} \over {1\over (p-2)}
\nu_c^{(3-p)/2}- \nu_m^{(3-p)/2}}, & \hbox{for $\nu_m<\hat{\nu}<\nu_c$;} \\
1-{(3-p)\nu_c^{1/2}\hat{\nu}^{(2-p)/2}
\over \nu_c^{(3-p)\over 2}-(p-2) \nu_m^{(3-p)\over 2}}, & \hbox{for $\nu_c<\hat{\nu}$.} \\
\end{array}%
\label{eq:Y_KN1} \right.
\end{equation}

For $1<p<2$, the total energy emitted is $\int F_\nu d\nu =
{2F_0\over (2-p)(3-p)}\nu_c^{(p-1)/2}S_1$, where $S_1=
[(3-p)\nu_c^{1/2}\nu_M^{(2-p)/2}- \nu_c^{(3-p)/2} -(2-p)
\nu_m^{(3-p)/2}]$. Now $\eta_{_{KN}}$ can be estimated as
\begin{equation} \eta_{_{KN}}\sim
\left\{%
\begin{array}{ll}
    0, & \hbox{for $\hat{\nu}<\nu_m$;} \\
    {(2-p)(\hat{\nu}^{(3-p)/2}-\nu_m^{(3-p)/2}) \over S_1}, & \hbox{for $\nu_m<\hat{\nu}<\nu_c$;} \\
1-{(3-p)\nu_c^{1\over 2}(\nu_M^{(2-p)\over
2}-\hat{\nu}^{(2-p)\over 2}) \over
S_1}, & \hbox{for $\nu_c<\hat{\nu}<\nu_M$.} \\
\end{array}%
\label{eq:Y_KN3} \right.
\end{equation}

{\bf B. Fast cooling.}

\begin{equation} F_\nu =F_0
\left\{%
\begin{array}{ll}
    (\nu/\nu_m)^{-1/2}, & \hbox{for $\nu_c<\nu<\nu_m$;} \\
    (\nu/\nu_m)^{-p/2}, & \hbox{for $\nu_m<\nu<\nu_M$.} \\
\end{array}%
\right.
\end{equation}
For $p>2$, the total energy emitted is $\int F_\nu d\nu =
2F_0\nu_m^{1/2} [({p-1\over p-2}) \nu_m^{1/2} - \nu_c^{1/2}]$,
where the emission below $\nu_c$ has been ignored. The
$\eta_{_{KN}}$ is estimated as
\begin{equation} \eta_{_{KN}}\sim
\left\{%
\begin{array}{ll}
    0, & \hbox{for $\hat{\nu}<\nu_c$;} \\
    {\hat{\nu}^{1/2} - \nu_c^{1/2} \over ({p-1\over p-2}) \nu_m^{1/2} - \nu_c^{1/2}},
    & \hbox{for $\nu_c<\hat{\nu}<\nu_m$;} \\
1-{\nu_m^{(p-1)/2}\hat{\nu}^{(2-p)/2}
\over ({p-1}) \nu_m^{1/2} - (p-2)\nu_c^{1/2}}, & \hbox{for $\nu_m<\hat{\nu}$.} \\
\end{array}%
\label{eq:Y_KN2} \right.
\end{equation}

For $1<p<2$, the total energy emitted is $\int F_\nu d\nu =
{2F_0\nu_m^{1/2} \over 2-p}S_2$, where $S_2=
[\nu_m^{(p-1)/2}\nu_M^{(2-p)/2}-(p-1) \nu_m^{1/2} -
(2-p)\nu_c^{1/2}]$. We have
\begin{equation} \eta_{_{KN}}\sim
\left\{%
\begin{array}{ll}
    0, & \hbox{for $\hat{\nu}<\nu_c$;} \\
{(2-p)(\hat{\nu}^{1/2} - \nu_c^{1/2}) \over S_2},
    & \hbox{for $\nu_c<\hat{\nu}<\nu_m$;} \\
1-{\nu_m^{(p-1)/2}[\nu_M^{(2-p)/2}-\hat{\nu}^{(2-p)/2}]
\over S_2}, & \hbox{for $\nu_m<\hat{\nu}<\nu_M$.} \\
\end{array}%
\label{eq:Y_KN4} \right.
\end{equation}

\section{When is the Klein-Nishina correction important?}\label{sec:Time}
In the shock front, the magnetic field strength $B$ is
\begin{equation}
B=0.04\Gamma \epsilon_{B,-2}^{1/2}n_0^{1/2}.
\end{equation}
The typical synchrotron radiation frequency of an electron with
random Lorentz factor $\gamma_e$ is
\begin{equation}
\nu(\gamma_e)={2.8\times 10^6\over 1+z}~{\rm Hz}~\gamma_e^2 \Gamma
B,
\end{equation}

{\bf A. The XRT lightcurve}\\
For $\nu_X\sim 10^{17}$ Hz, we have $\gamma_e(\nu_X) = 1.3\times
10^5[2/(1+z)]^{-1/2}
\Gamma_1^{-1}\epsilon_{B,-2}^{-1/4}n_0^{-1/4}$ and
\begin{equation}
\hat{\nu} \sim 5\times 10^{15} {\rm
Hz}~[(1+z)/2]^{1/2}\Gamma_1^2\epsilon_{B,-2}^{1/4}n_0^{1/4},
\end{equation}

Therefore, the Klein-Nishina correction seems to be unimportant
(i.e., $\eta_{_{KN}}\sim 1$) for $t_d \sim 1$ (when $\hat{\nu}
\sim \nu_c$) and $\epsilon_B\sim 0.01$.

{\bf B. The R-band lightcurve}\\
For $\nu_R\sim 4.3\times 10^{14}$ Hz, we have $\gamma_e(\nu_R) =
8\times 10^3[2/(1+z)]^{-1/2}
\Gamma_1^{-1}\epsilon_{B,-2}^{-1/4}n_0^{-1/4}$ and
\begin{equation}
\hat{\nu} \sim 8\times 10^{16} {\rm
Hz}~[(1+z)/2]^{1/2}\Gamma_1^2\epsilon_{B,-2}^{1/4}n_0^{1/4},
\end{equation}

Then, with $\epsilon_B\sim 0.01$, the Klein-Nishina correction
seems to be unimportant for a long time. On the other hand, the
factor $\eta\simeq \min\{ 1, (\nu_m/\nu_c)^{(p-1)/2}\}\sim 1$ for
$t_d<1$. As a consequence, the inverse Compton effect is very
important both for the long wavelength afterglow calculation and
for the X-ray lightcurve calculation. However, it may be
unimportant for a lower $\epsilon_B$ since $\nu_c\propto
\epsilon_B^{-3/2}$.

\end{appendix}

\end{document}